\documentclass[journal, 10pt]{IEEEtran}
\usepackage{amsmath,graphicx}
\usepackage{pifont}
\newcommand{\cmark}{\ding{51}}%
\newcommand{\xmark}{\ding{53}}%
\usepackage{amssymb}
\usepackage{bm}
\usepackage{multirow}
\usepackage{adjustbox}
\usepackage{tabularx}
\usepackage{tikz}
\usepackage{caption}
\usepackage{subcaption}
\usepackage{verbatim}
\usepackage{float}
\usepackage{commath}
\usepackage{url}
\usepackage{soul}
\usepackage{colortbl}
\newcommand{\floor}[1]{\left\lfloor #1 \right\rfloor}
\newcommand{\ceil}[1]{\left\lceil #1 \right\rceil}

\begin{document}

\title{Dense CNN with Self-Attention for Time-Domain Speech Enhancement }

\author{Ashutosh~Pandey, ~\IEEEmembership{Student~Member,~IEEE}~and~DeLiang~Wang,~\IEEEmembership{Fellow,~IEEE}

\thanks{This research was supported in part by two NIDCD grants (R01DC012048 and R02DC015521) and the Ohio Supercomputer
Center.}
\thanks{A. Pandey is with the Department of Computer Science and Engineering, The Ohio State University, Columbus, OH 43210 USA (e-mail:
pandey.99@osu.edu).}
\thanks{D. L. Wang is with the Department of Computer Science and Engineering
and the Center for Cognitive and Brain Sciences, The Ohio State University,
Columbus, OH 43210 USA (e-mail: dwang@cse.ohio-state.edu)}\vspace{-0.5em}}

\maketitle

\begin{abstract}
Speech enhancement in the time domain is becoming increasingly popular in recent years, due to its capability to jointly enhance both the magnitude and the phase of speech. In this work, we propose a dense convolutional network (DCN) with self-attention for speech enhancement in the time domain. DCN is an encoder and decoder based architecture with skip connections. Each layer in the encoder and the decoder comprises a dense block and an attention module. Dense  blocks and attention modules help in feature extraction using a combination of feature reuse, increased network depth, and maximum context aggregation. Furthermore, we reveal previously unknown problems with a loss based on the spectral magnitude of enhanced speech. To alleviate these problems, we propose a novel loss based on magnitudes of enhanced speech and a predicted noise. Even though the proposed loss is based on magnitudes only, a constraint imposed by noise prediction ensures that the loss enhances both magnitude and phase. Experimental results demonstrate that DCN trained with the proposed loss substantially outperforms other state-of-the-art approaches to causal and non-causal speech enhancement.
 \end{abstract}


\IEEEpeerreviewmaketitle

\section{Introduction}
Speech signal in a real-world environment is degraded by background noise that reduces its intelligibility and quality for human listeners. Further, it can severely degrade the performance of speech-based applications, such as automatic speech recognition (ASR), teleconferencing, and hearing-aids. Speech enhancement aims at improving the intelligibility and quality of a speech signal by removing or attenuating background noise. It is used as preprocessor in speech-based applications to improve their performance in noisy environments. Monaural (single-channel) speech enhancement provides a versatile and cost-effective approach to the problem by utilizing recordings from a single microphone. Single-channel speech enhancement in low signal-to-noise ratio (SNR) conditions is considered a very challenging problem. This study focuses on single-channel speech enhancement in the time domain.

Traditional monaural speech enhancement approaches include spectral subtraction, Wiener filtering and statistical model-based methods \cite{loizou2013speech}. Speech enhancement has been extensively studied in recent years as a supervised learning problem using deep neural networks (DNNs) since the first study in \cite{wang2013towards}.

Supervised approaches to speech enhancement generally convert a speech signal to a time-frequency (T-F) representation, and extract input features and training targets from it \cite{wang2017supervised}. Training targets are either masking based or mapping based \cite{wang2014training}. Masking based targets, such as the ideal ratio mask (IRM) \cite{wang2014training} and phase sensitive mask \cite{erdogan2015phase}, are based on time-frequency relation between noisy and clean speech, whereas mapping based targets \cite{lu2013speech, xu2015regression}, such as spectral magnitude and log power spectrum, are based on clean speech. Input features and training targets are used to train a DNN that estimates targets from noisy features. Finally, enhanced waveform is obtained by reconstructing a signal from the estimated target.

Most of the T-F representation based methods aim to enhance only spectral magnitudes and noisy phase is used unaltered for time-domain signal reconstruction \cite{lu2013speech, xu2015regression, weninger2015speech, chen2016large, fu2016snr, park2016fully, chen2017long, tan2018gated1}. This is mainly because phase was considered not important for speech enhancement \cite{wang1982unimportance}, and exhibits no spectro-temporal structure amenable to supervised learning \cite{williamson2016complex}. A recent study, however, found that the phase can play an important role in the quality of enhanced speech, especially in low SNR conditions \cite{paliwal2011importance}. This has led researchers to explore techniques to jointly enhance magnitude and phase \cite{williamson2016complex, fu2017complex, pandey2019exploring, tan2019learning}. 

There are two approaches to jointly enhance magnitude and phase: complex spectrogram enhancement and time-domain enhancement. In complex spectrogram enhancement, the real and the imaginary part of the complex-valued noisy STFT (short-time Fourier transform) is enhanced. Based on training targets, complex spectrogram enhancement is further categorized as complex ratio masking \cite{williamson2016complex} and complex spectral mapping \cite{fu2017complex, pandey2019exploring, tan2019learning}.

Time-domain enhancement aims at directly predicting enhanced speech samples from noisy speech samples, and in the process, magnitude and phase are jointly enhanced \cite{pascual2017segan, rethage2017wavenet, qian2017speech, fu2018end, pandey2019new, pandey2019tcnn, pandey2020densely}. Even though complex spectrogram enhancement and time-domain enhancement have similar objectives, time-domain enhancement has some advantages. First, time-domain enhancement avoids the computations associated with the conversion of a signal to and from the frequency domain. Second, since the underlying DNN is trained from raw samples, it can potentially learn to extract better features that are suited for the particular task of speech enhancement. Finally, short-time processing based on a T-F representation requires frame size to be greater than some threshold to have sufficient spectral resolution, whereas in time-domain processing frame size can be set to an arbitrary value. In \cite{luo2019conv} and \cite{luo2020dual}, the performance of a time-domain speaker separation network is substantially improved by setting frame size to very small values. However, using a smaller frame size requires more computations due to an increased number of frames.

Self-attention is a widely utilized mechanism for sequence-to-sequence tasks, such as machine translation \cite{vaswani2017attention}, image generation \cite{zhang2019self} and ASR \cite{dong2018speech}. First introduced in \cite{vaswani2017attention}, self-attention is a mechanism for selective context aggregation, where a given output in a sequence is computed based on only a subset of the input sequence (attending on that subset) that is helpful for the output prediction. It can be utilized for any task that has sequential input and output. Self-attention can be a helpful mechanism for speech enhancement because of the following reason. A spoken utterance generally contains many repeating phones. In a low SNR condition, a given phone can be present in both high and low SNR regions in the utterance. This suggests that a speech enhancement system based on self-attention can attend over phones in high SNR regions to better reconstruct phones in low SNR regions. Recent studies \cite{zhao2020monaural}, \cite{giri2019attention}, \cite{kim2020t}, and \cite{koizumi2020speech} have successfully employed self-attention for speech enhancement with promising results. 

In this work, we propose a dense convolutional network (DCN) with self-attention for speech enhancement in the time domain. DCN is based on an encoder-decoder architecture with skip connections \cite{pandey2019new, pandey2019tcnn, pandey2020densely}. Each of the layers in the encoder and the decoder comprises a dense block \cite{huang2017densely} and an attention module. The dense block is used for better feature extraction with feature reuse in a deeper network, and the attention module is used for utterance level context aggregation. This study is an extension of our previous work in \cite{pandey2020densely}, where dilated convolutions are utilized inside a dense block for context aggregation. We find attention to be superior to dilated convolutions for speech enhancement. We use an attention module similar to the one proposed in \cite{liu2020voice}. 

Furthermore, we find that the spectral magnitude (SM) loss  proposed for training of a time-domain network \cite{Pandey2018} obtains better objective intelligibility and quality scores, but introduces a previously unknown artifact in enhanced utterances.  Also, it is inconsistent in terms of SNR improvement. We propose a magnitude based loss to remove this artifact and obtain consistent SNR improvement as a result.  The proposed loss function is based on spectral magnitudes  of the enhanced speech and a predicted noise. In case of perfect estimation, the proposed loss reduces the possible number of phase values at a given T-F unit from infinity to two, one of which corresponds to the clean phase, i.e, it constrains the phase to be much closer to clean phase. We call this loss phase constrained magnitude (PCM) loss. 

The rest of the paper is organized as follows. We describe speech enhancement in the time domain in Section \ref{sec_definition}. DCN architecture and its building blocks are explained in Section \ref{sec_dcn}. Section \ref{sec_loss} describes different loss functions along with the proposed loss. Experimental settings are given in Section \ref{sec_experiments}, and results are discussed in Section \ref{sec_results}. Concluding remarks are given in Section \ref{sec_conclusions}. 

\section{Speech Enhancement in the Time Domain}
\label{sec_definition}
Given a clean speech signal $\bm{s}$ and a noise signal $\bm{n}$, the noisy speech signal is modeled as
\begin{equation}
\bm{y} = \bm{s} + \bm{n}
\end{equation}
where \{$\bm{y}$, $\bm{s}$, $\bm{n}$\} $ \in \mathbb{R}^{M \times 1}$, and $M$ represents the number of samples in the signal. The goal of a speech enhancement algorithm is to get a close estimate, $\widehat{\bm{s}}$,  of $\bm{s}$ given $\bm{y}$.

Speech enhancement in the time domain aims at computing  $\widehat{\bm{s}}$ directly from $\bm{y}$ instead of using a T-F representation of  $\bm{y}$. We can formulate time-domain enhancement using a DNN as 
\begin{equation}
\label{eq_dnn_sig}
\widehat{\bm{s}}= f_{\bm{\theta}}(\bm{y})
\end{equation} 
where $f_{\bm{\theta}}$ denotes a function defining a DNN model parametrized by $\bm{\theta}$. The DNN model $f_{\bm{\theta}}$ can be any of the existing DDN architectures such as a feedforward, recurrent, or convolutional neural network.

\subsection{Frame-Level Processing}
Generally, the input signal $\bm{y}$ is first chunked into overlapping frames which is then processed as frame-level enhancement. Let $\bm{Y} \in \mathbb{R}^{T \times L} $ denote the matrix containing frames of signal $\bm{y}$, and $\bm{y}_{t} \in \mathbb{R}^{L \times 1}$ the $t^{th}$ frame.  $\bm{y}_{t}$ is defined as
\begin{equation}
y_{t}[k] = y[(t-1)\cdot J + k], \  k = 0, \cdots, L-1
\end{equation}
where $T$ is the number of frames, $L$ is the frame length, and $J$ is the frame shift. $T$ is given by $\ceil{\frac{M}{J}}$, where $\ceil{\ }$ denotes the ceiling function. Note that $\bm{y}$ is padded with zeros if $M$ is not divisible by $J$. Frame-level processing using a DNN can be defined as
\begin{equation}
\label{eq_dnn_frame}
\widehat{\bm{s}}_{t} = f_{\bm{\theta}}(\bm{y}_{t-K_{1}}, \cdots, \bm{y}_{t-1}, \bm{y}_{t}, \bm{y_{t+1}}, \cdots, \bm{y}_{t+K_{2}})
\end{equation}
where $\widehat{\bm{s}}_{t}$ is computed using $\bm{y}_{t}$, $K_{1}$ past frames, and $K_{2}$ future frames.  

\subsection{Causal Speech Enhancement}

A speech enhancement system is considered causal if the prediction for a given frame is computed using only the current and the past frames. This can be defined as
\begin{equation}
\widehat{\bm{s}}_{t} = f_{\bm{\theta}}(\bm{y}_{t-K_{1}}, \cdots, \bm{y}_{t-1}, \bm{y}_{t})
\end{equation}
A causal speech enhancement system is required for real-time speech enhancement.
\begin{figure*}[!t]
\centering
\includegraphics[width=0.8\textwidth]{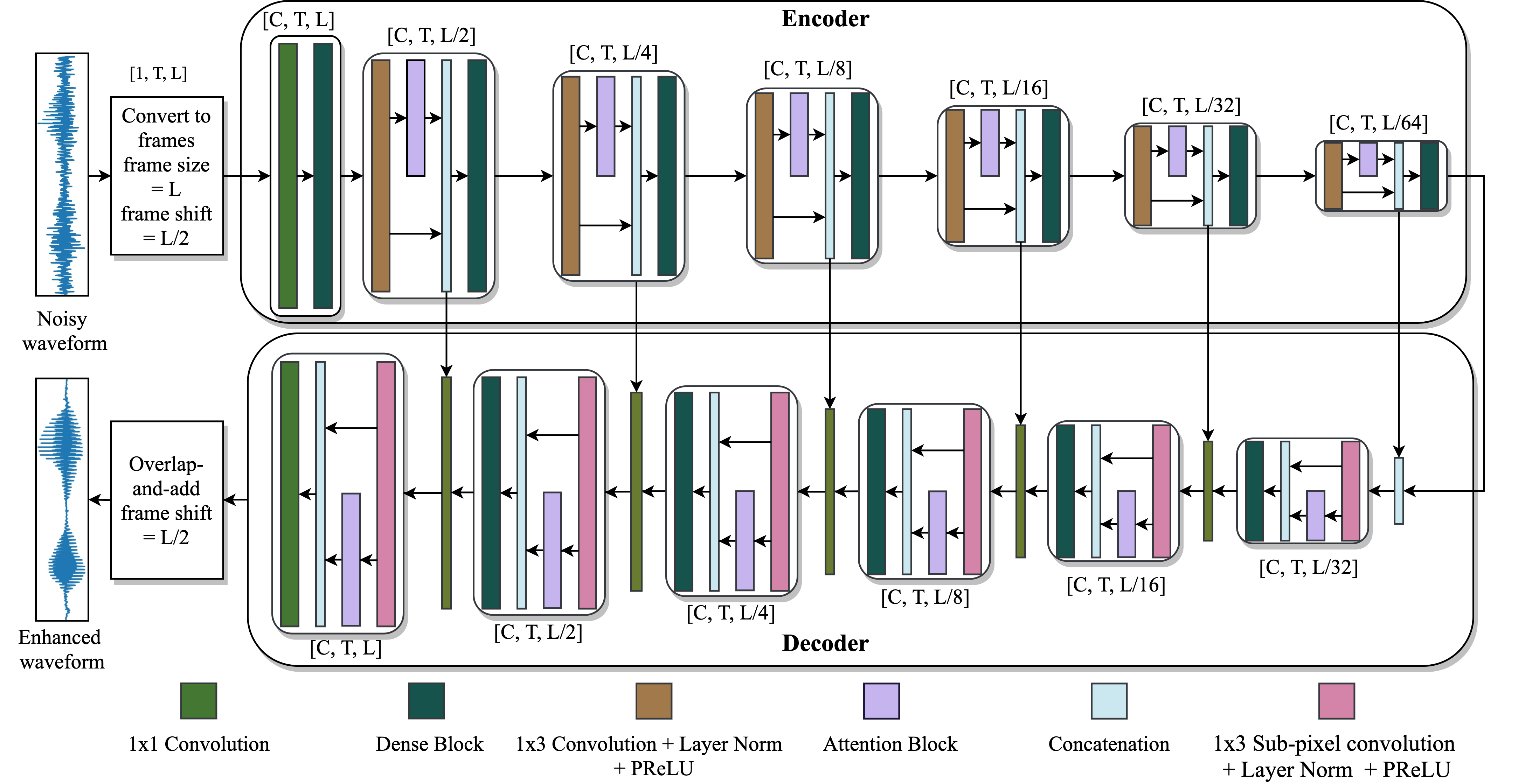}
\caption{Diagram of the proposed DCN model.}
\label{fig_sesan}
\end{figure*}

\section{Dense Convolutional Network}
\label{sec_dcn}
A block diagram of DCN is shown in Fig. \ref{fig_sesan}. The building blocks of DCN are 2D convolution, sub-pixel convolution \cite{shi2016real}, layer normalization \cite{{ba2016layer}}, dense block \cite{huang2017densely}, and self-attention module \cite{vaswani2017attention}. Next, we describe these building blocks one by one.   

\subsection{2-D Convolution}
\begin{figure}[!b]
\centering
\includegraphics[width=0.45\textwidth, keepaspectratio]{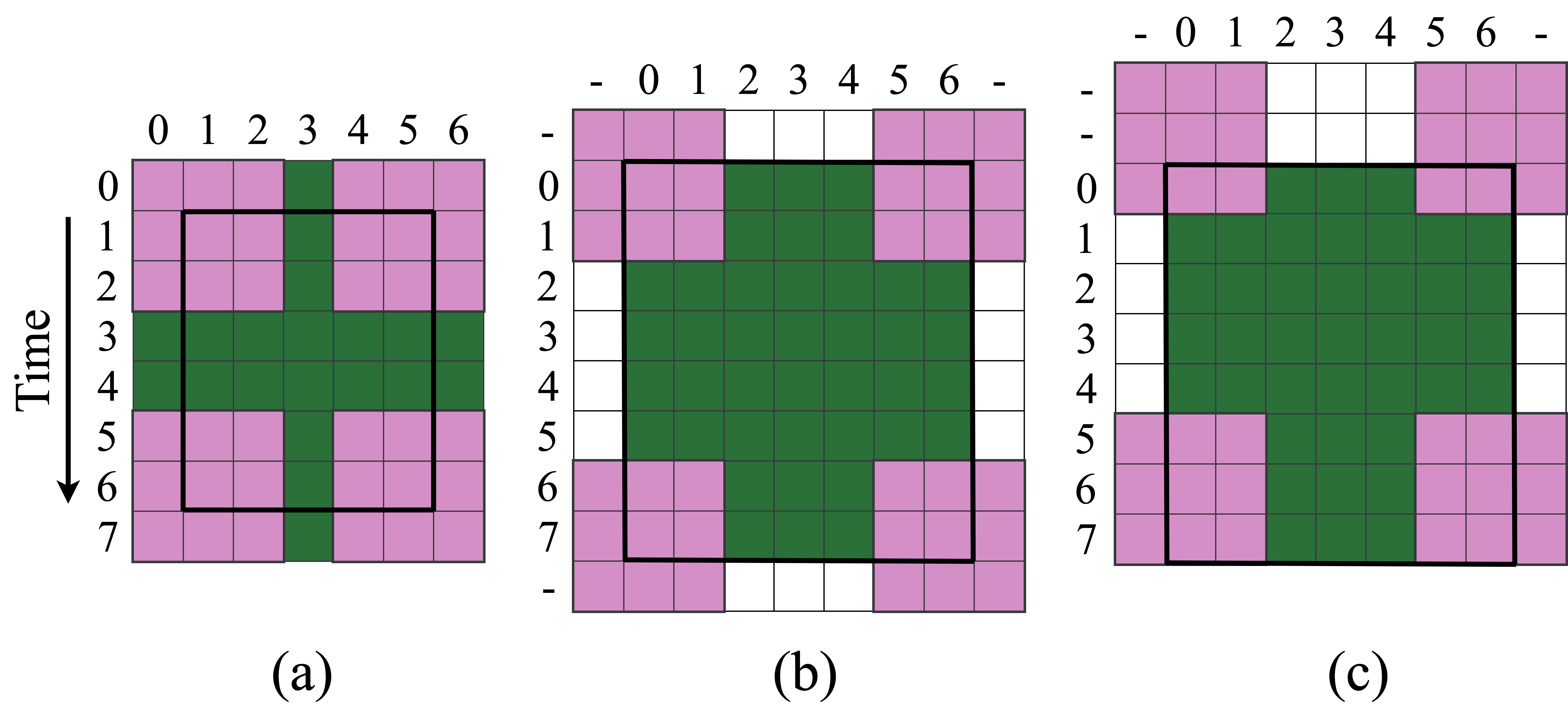}
\caption{Illustration of different types of convolution of an input of size $8 \times 7$ with a kernel of size $3 \times 3$. (a) VALID convolution, (b) Non-causal convolution with SAME padding. (c) Causal convolution along time with SAME padding.}
\label{fig_conv}
\end{figure}

Formally, a 2-D discrete convolution operator $*$, which convolves a signal $\bm{Y}$ of size $T \times L$ with a kernel $\bm{K}$ of size $m \times n$ and stride $(r, s)$, is defined as 
\begin{equation}
\label{eq_conv}
(\bm{Y}*\bm{K})(i, j) = \sum_{u = 0}^{m -1}\sum_{v=0}^{n-1}Y(r \cdot i + u, s \cdot j + v) \cdot K(u, v)
\end{equation}
where $i \in \{0, 1 \cdots, T - m\}$ and $j \in \{0, 1, \cdots L-n\}$. Note that Eq. (\ref{eq_conv}) is actually a correlation operator generally referred as convolution in convolutional neural networks. Further, Eq. (\ref{eq_conv}) defines VALID convolution in which the kernel is placed only at the locations where it does not cross the signal boundary, and as a result the output size is reduced to $(T-m+1) \times (L-n+1)$. Fig. \ref{fig_conv}(a) illustrates the position of kernel on four corners for VALID convolution. To obtain an output of the same size as the input, the input is padded with zeros around all the boundaries, and is known as SAME padding, which is shown in Fig. \ref{fig_conv}(b).

Causal convolution is a term used for convolution with time-series signals, such as audio and video. A convolution is considered causal if the output at $t$ is computed using inputs at time instances less than or equal to $t$. For speech enhancement, the matrix $\bm{Y}$, which stores the frames of speech signal, $\bm{y}_{0}, \bm{y}_{1}, \cdots, \bm{y}_{t}, \cdots, \bm{y}_{T-1}$, is a time series. A non-causal convolution can be easily converted to a causal one by padding extra zeros in the beginning ($t<0$). A causal convolution is shown in Fig. \ref{fig_conv}(c). In general, a padding of length $m-1$ is required for causal convolution with a kernel of size $m$ along the time dimension.   
\begin{figure}[!t]
\centering
\includegraphics[width=0.45\textwidth]{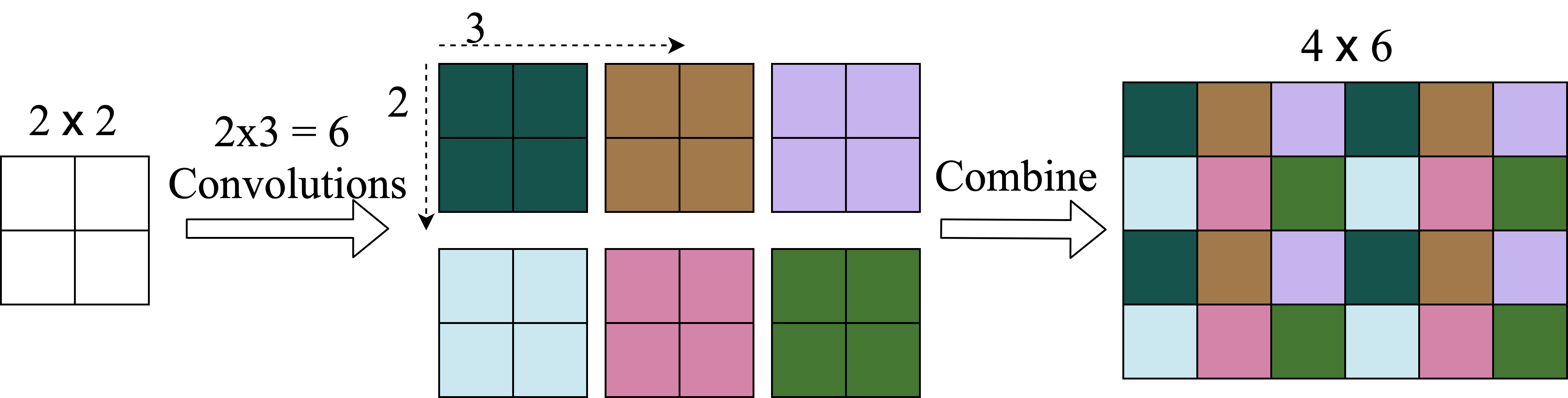}
\caption{An illustration of sub-pixel convolution for upsampling a 2D signal by rate $(2, 3)$.}
\label{fig_sub_pixel}
\end{figure}

\subsection{Sub-pixel Convolution}
First proposed in \cite{shi2016real}, a sub-pixel convolution is used to increase the size of a signal (upsampling). It becomes increasingly popular as an alternative to transposed convolution, as it avoids a well-known checkerboard artifact in the output signal \cite{odena2016deconvolution} and is computationally efficient. For an upsampling rate $(r, s)$, sub-pixel convolution uses $r \cdot s$ convolutions to obtain $r \cdot s $ different signals of the same size as the input. The different convolutions in a sub-pixel convolution are defined as 
\begin{equation}
\begin{aligned}
\bm{S}_{0, 0} &= \text{Pad}(\bm{Y}) * \bm{K}_{0, 0} \\
\bm{S}_{0, 1} &= \text{Pad}(\bm{Y}) * \bm{K}_{0, 1} \\
& \cdots \\
\bm{S}_{r-1, s-1} &= \text{Pad}(\bm{Y}) * \bm{K}_{r-1, s-1} 
\end{aligned}
\end{equation}
where Pad denotes the SAME padding operation and $\bm{K}_{i, j}$ denotes a convolution kernel. $\bm{S_{1, 1}}$, $\bm{S_{1, 2}}$, $\cdots$,  and $\bm{S_{r-1, s-1}}$ are combined to obtain the upsampled signal using the following equation,
\begin{equation}
S(i, j) = S_{(i\%r), (j\%s)}(\floor{i/r}, \floor{j/s})
\end{equation}
where $\%$ denotes the remainder operator, $\floor{\ }$ the floor operator, $i \in \{0, 1, \cdots, r \cdot T -1\}$, and $j \in \{0, 1, \cdots, s \cdot L -1\}$. A diagram of sub-pixel convolution is shown in Fig. \ref{fig_sub_pixel}. 

\subsection{Layer Normalization}
Layer normalization is a technique proposed to improve generalization and facilitate DNN training \cite{ba2016layer}. It is used as an alternative to batch normalization, which is sensitive to training batch size.  We use the following layer normalization.
\begin{equation}
\bm{y}^{norm} = \frac{\bm{y} - \mu_{y}}{\sqrt{\sigma^{2}_{y} + \epsilon}} \odot \bm{\gamma} + \bm{\beta}
\end{equation}
where $ \mu_{y}$ and $\sigma_{y}^{2}$, respectively, are scalars representing mean and variance of $\bm{y}$. $\bm{\gamma}$ and $ \bm{\beta}$ are trainable variables of the same size as $\bm{y}$, and $-, +$, and $\odot$ respectively denote element-wise subtraction, addition and multiplication. $\epsilon$ is a small positive constant to avoid division by zero. For an input
of shape $[C, T, L]$ ($C$ channels, $T$ frames), normalization is performed over the last dimension using $\bm{\gamma}$ and $\bm{\beta}$ that are shared across channels and frames.


\subsection{Dense Block}
\begin{figure}[!h]
\centering
\includegraphics[width=0.48\textwidth, keepaspectratio]{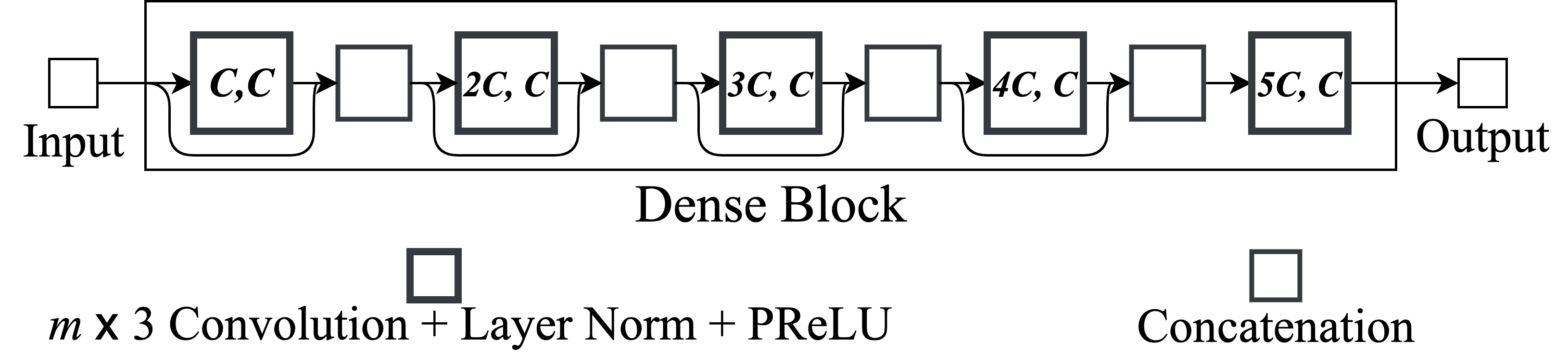}
\caption{The proposed dense block. $X$ and $Y$ in the pair $(X, Y)$ inside convolution box, respectively, denote the number of input and output channels.}
\label{fig_dense}
\end{figure}

Densely connected convolutional networks were recently proposed in \cite{huang2017densely}. A densely connected network is based on the idea of feature reuse in which an output at a given layer is reused multiple times in the subsequent layers. In other words, the input to a given layer is not just the input from the previous layer but also the outputs from several layers before the given layer. It has two major advantages. First, it can avoid the vanishing gradient problem in DNNs because of the direct connections of a given layer to the subsequent layers. Second, a thinner (in terms of the number of channels) dense network is found to outperform a wider normal network, and hence improves the parameter efficiency of the network. Formally, a dense connection can be defined as
\begin{equation}
\bm{y}^{l}= g(\bm{y}^{l-1},\bm{y}^{l-2}, \cdots, \bm{y}^{l-D})
\end{equation}
where $\bm{y}^{l}$ denotes the output at layer $l$, $g$ is the function represented by a single layer in the network, and $D$ is the depth of dense connections. DCN uses a dense block after each layer in the encoder and the decoder. The proposed dense block is shown in Fig. \ref{fig_dense}. It consists of five convolutional layers with $m \times 3$ convolutions followed by layer normalization and parametric ReLU nonlinearity \cite{he2015delving}. We set $m$ to $2$ for causal and to $3$ for non-causal convolution. The input to a given layer is formed by a concatenation of the input to and the output of the previous layer. The number of input channels in the successive layers increases linearly as $C, 2C, 3C, 4C, 5C$. The output after each convolution has $C$ channels.

\subsection{Self-attention Module}
\label{sec_attention}
DCN uses self attention after downsampling in the encoder and upsampling in the decoder. An attention mechanism comprises three key components: query $\bm{Q}$, key $\bm{K}$, and value $\bm{V}$, where $\{\bm{K}, \bm{Q}\} \in \mathbb{R}^{T \times I}$ and $\bm{V} \in \mathbb{R}^{T \times J}$. First, correlation scores of all the rows in $\bm{Q}$ are computed with all the rows in $\bm{K}$ using the following equation.  
\begin{equation}
 \bm{W} = \bm{Q}\bm{K}^{\mathbb{T}}
\end{equation}
where $\bm{K}^{\mathbb{T}}$ denotes the transpose of $\bm{K}$ and $\bm{W} \in \mathbb{R}^{T \times T}$. Next, correlation scores are converted to probability values using a Softmax operation defined as 
\begin{equation}
\text{Softmax}(\bm{W})(i, j) = \frac{\exp^{W(i,j)}}{\sum_{j=0}^{T-1}\exp^{W(i, j)}}
\end{equation}
Finally, the rows of $\bm{V}$ are linearly combined using weights in $\text{Softmax}(\bm{W})$ to obtain the attention output.
\begin{equation}
\bm{A} = \text{Softmax}(\bm{W})\bm{V}
\end{equation}
An attention mechanism is called self-attention if $\bm{Q}$ and $\bm{K}$ are computed from the same sequence. For example, given an input sequence $\bm{Y}$, a self-attention layer can be implemented by using a linear layer to compute $\bm{Q}, \bm{K},$ and $\bm{V}$, and then using Eqs. (11-13) to get the attention output. 

\begin{figure}[!b]
\centering
\includegraphics[width=0.5\textwidth]{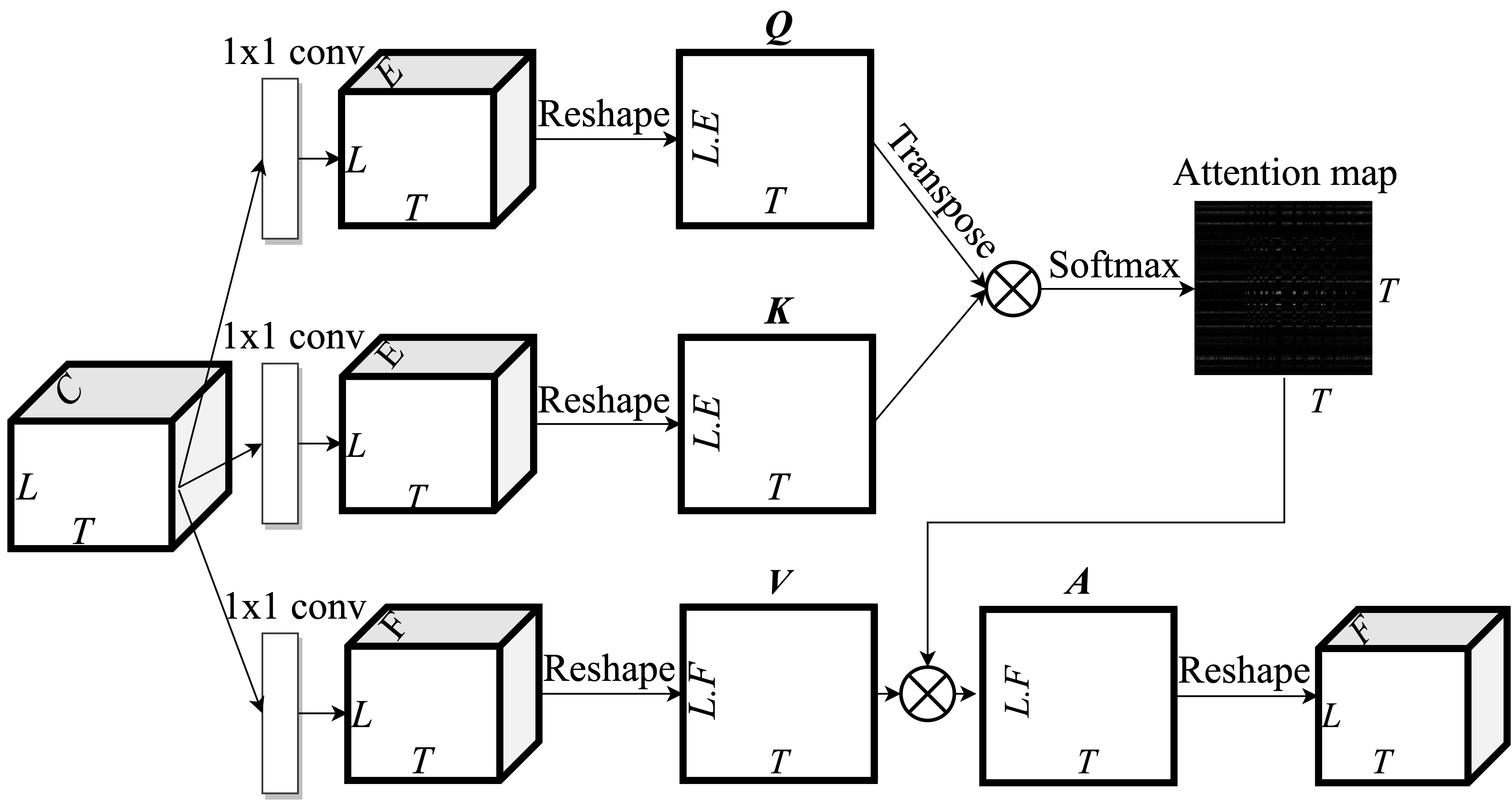}
\caption{Proposed self-attention module.}
\label{fig_attn}
\end{figure}

The proposed self-attention module in DCN is shown in Fig. \ref{fig_attn}. First, three different $1 \times 1$ convolutions are used to transform an input of shape $[C, T, L] $ to $\bm{Q}$ of shape $[E, T, L]$, $\bm{K}$ of shape  $[E, T, L]$, and $\bm{V}$ of shape $[F, T, L]$. Next, $\bm{Q}, \bm{K}$, and $\bm{V}$ are reshaped to obtain 2D matrices. Finally, Eq. (11), Eq. (12), and Eq. (13) are applied to get the 2D attention output, which is reshaped to get an output of shape $[F, T, L]$. The proposed attention module is similar to the one in \cite{liu2020voice} with one difference: we do not use linear layers to project $\bm{Q}$ and $\bm{K}$ to lower dimensions. We find that the performance is similar with and without linear layers.

Causal attention can be implemented by applying a mask to $\bm{W}$ where entries above the main diagonal are set to negative infinity so that the contribution from future frames in Eq. (12) becomes zero. This can be defined as  
\begin{equation}
\bm{A}_{causal} = \text{Softmax}(\text{Mask}(\bm{W}))\bm{V}
\end{equation}
where
\begin{equation}
	\text{Mask}(W)(i, j) = \begin{cases}
	      W(i, j), & \text{if}\ i \leq j \\
	      -\infty, & \text{otherwise}
	    \end{cases}
\end{equation}

With the building blocks described, we now present the processing flow of DCN. First, a given utterance $\bm{y}$ is chunked into frames of size $L$, reshaped to a shape of $[1, T, L]$, and fed to the encoder. The first layer in the encoder uses $1 \times 1$ convolution to increase the number of channels to $C$, and then is processed by a dense block. The following $6$ layers in the encoder process their input by one convolutional layer for downsampling, one attention module and one dense block. The output of the attention module is concatenated with its input along the channel dimension before feeding it to the dense block. The output of the encoder is fed to the decoder. Each layer in the decoder has one module for upsampling using sub-pixel convolution, one attention module and one dense block. The output of the decoder is concatenated with the output of the corresponding symmetric layer in the encoder. The final layer in the decoder does not include a dense block, and uses $1 \times 1$ convolution to output a signal with 1 channel, which is subject to overlap-and-add to obtain the enhanced utterance. Each convolution in DCN, except at the input and at the output, is followed by layer normalization and parametric ReLU \cite{he2015delving}.

\section{Loss Functions}
\label{sec_loss}
\subsection{Time-Domain Loss}
An utterance level mean squared error (MSE) loss in the time domain is defined as 
\begin{equation}
L_{T}(\bm{s}, \bm{\widehat{s}}) = \frac{1}{M}\sum_{k=0}^{M-1}(s[k] - \widehat{s}[k])^{2} 
\end{equation}

\begin{figure*}[t!]
\centering
\includegraphics[width=0.8\textwidth]{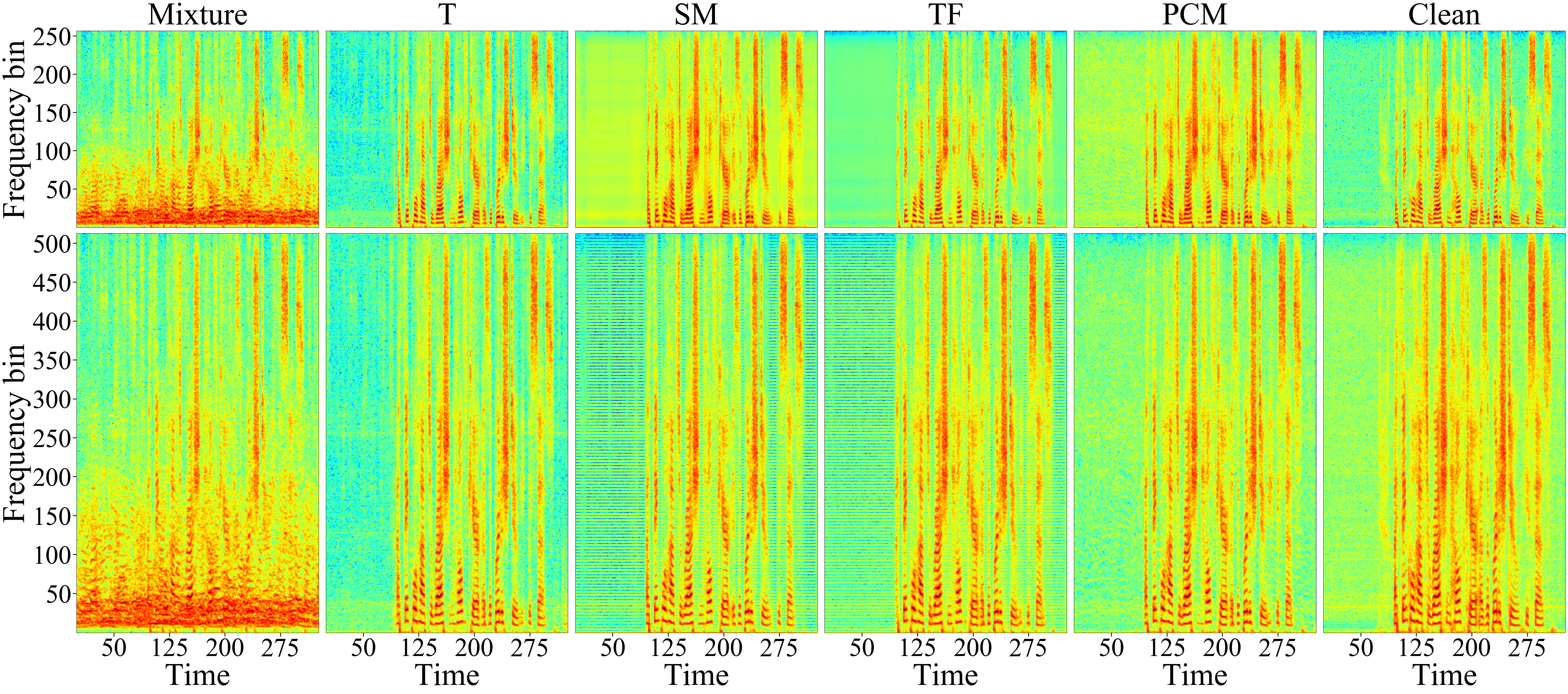}
\caption{Spectrograms of a sample utterance processed using DCN trained with different loss functions. Frame size for STFT is $32$ ms in the first row and $64$ ms in the second row. }
\label{fig_artifact}
\end{figure*}

\subsection{STFT Magnitude Loss}
\label{sec_lsm}
A loss based on STFT magnitude was proposed in \cite{pandey2019new}, which was found to be superior to the time-domain loss in terms of objective intelligibility and quality scores, and a little worse in terms of scale-invariant speech-to-distortion ratio (SI-SDR). The loss is defined as 
\begin{equation}
\begin{aligned}
L_{SM}(\bm{s}, \bm{\widehat{s}}) = \frac{1}{T\cdot F}\sum_{t=0}^{T-1}\sum_{f=0}^{F-1}&[(|S_{r}(t, f)| + |S_{i}(t,f)|)\\-&\ (|\widehat{S}_{r}(t,f)| + |\widehat{S}_{i}(t, f)|)] 
\end{aligned}
\end{equation}
where $\bm{S}$ and $\bm{\widehat{S}}$ respectively denote STFTs of $\bm{s}$ and $\bm{\widehat{s}}$, $T$ is the number of time frames, and $F$ is the number of frequency bins. Subscripts $r$ and $i$ respectively denote the real and the imaginary part of a complex variable. $L_{SM}$ is a mean absolute error loss between the $L_{1}$ norm of clean and estimated STFT coefficients \cite{pandey2018adversarial}.

Even though $L_{SM}$ can obtain better objective scores, it has some disadvantages. First, we find that it is not consistent in terms of SNR improvement, as in some cases processed SNR is found to be worse than unprocessed SNR. However, a consistent improvement is observed in scale-invariant scores, such as SI-SNR and SI-SDR, suggesting that enhanced utterances do not have an appropriate scale using $L_{SM}$, which is a requirement for speech enhancement algorithms. Second, we find that $L_{SM}$ introduces an unknown artifact in enhanced utterances, which does not affect intelligibility and quality scores, but this steady buzzing sound is annoying to human listeners. 

We find that the introduced artifact is not visible in a spectrogram with the same frequency resolution as in the STFT of $L_{SM}$. However, it can be observed with a higher frequency resolution. Spectrograms of a sample noisy utterance enhanced using DCN trained with different loss functions is plotted in Fig. \ref{fig_artifact}. The first row plots spectrograms with frame size and frame shift equal to the ones used in computation of $L_{SM}$, $L_{TF}$ (Eq. (18)), and $L_{PCM}$ (Eq. (19)). The second row plots spectrograms with a frame size twice that in the first row. We can see horizontal stripes in the second plot of $L_{SM}$ and $L_{TF}$, which are not visible in the first row, and these stripes correspond to the artifact in enhanced utterances. This artifact is not present with the time-domain MSE loss or PCM loss proposed in the study.

\subsection{Time-frequency Loss}
Time-frequency loss, which was proposed in \cite{pandey2020densely}, is a combination of $L_{T}$ and $L_{SM}$. It is defined as 
\begin{equation}
L_{TF} = \alpha \cdot L_{T} + (1-\alpha) \cdot L_{SM}
\label{eq_loss_tf}
\end{equation}
where $\alpha$ is a hyperparameter. We find that $L_{TF}$ can solve the inconsistent SNR problem associated with $L_{SM}$ as it obtains consistent SNR improvement similar to $L_{T}$. Additionally, $L_{TF}$ preserves improvements in objective scores obtained using $L_{SM}$. However, $L_{TF}$ is not able to remove the artifacts, as shown in Fig. 6. We have explored different values of $\alpha$ in Eq. (\ref{eq_loss_tf}) and find that the artifact is present for a wide range of $\alpha$ values, and not straightforward to find a value that can remove the artifacts while maintaining objective scores similar to $L_{SM}$.

\subsection{Phase Constrained Magnitude Loss}
\label{sec_pcm}
We propose a new loss that is based on STFT magnitude but can alleviate both the problems associated with $L_{SM}$. Given $\bm{y}$, $\bm{s}$, and $\bm{\widehat{s}}$, a prediction for noise can be defined as
\begin{equation}
\bm{\widehat{n}} = \bm{y} - \bm{\widehat{s}}
\end{equation}
Now, we can modify the objective of speech enhancement to match not only the STFT magnitude of speech but that of the noise also. The PCM loss is defined as 
\begin{equation}
\begin{aligned}
L_{PCM}(\bm{s}, \bm{\widehat{s}}) &= \frac{1}{2}\cdot L_{SM}(\bm{s}, \bm{\widehat{s}}) + \frac{1}{2} \cdot L_{SM}(\bm{n}, \bm{\widehat{n}})
\end{aligned}
\end{equation}
Even though one can play with relative contributions of speech and noise, we find that the equal contribution in Eq. (20) obtains consistent SNR improvement similar to $L_{T}$, removes artifacts associated with $L_{SM}$, and achieves objective intelligibility and quality scores similar to $L_{SM}$.

\begin{figure}[!h]
\label{fig_pcm}
\centering
\begin{subfigure}[h]{0.22\textwidth}
\centering
\includegraphics[width=0.98\textwidth]{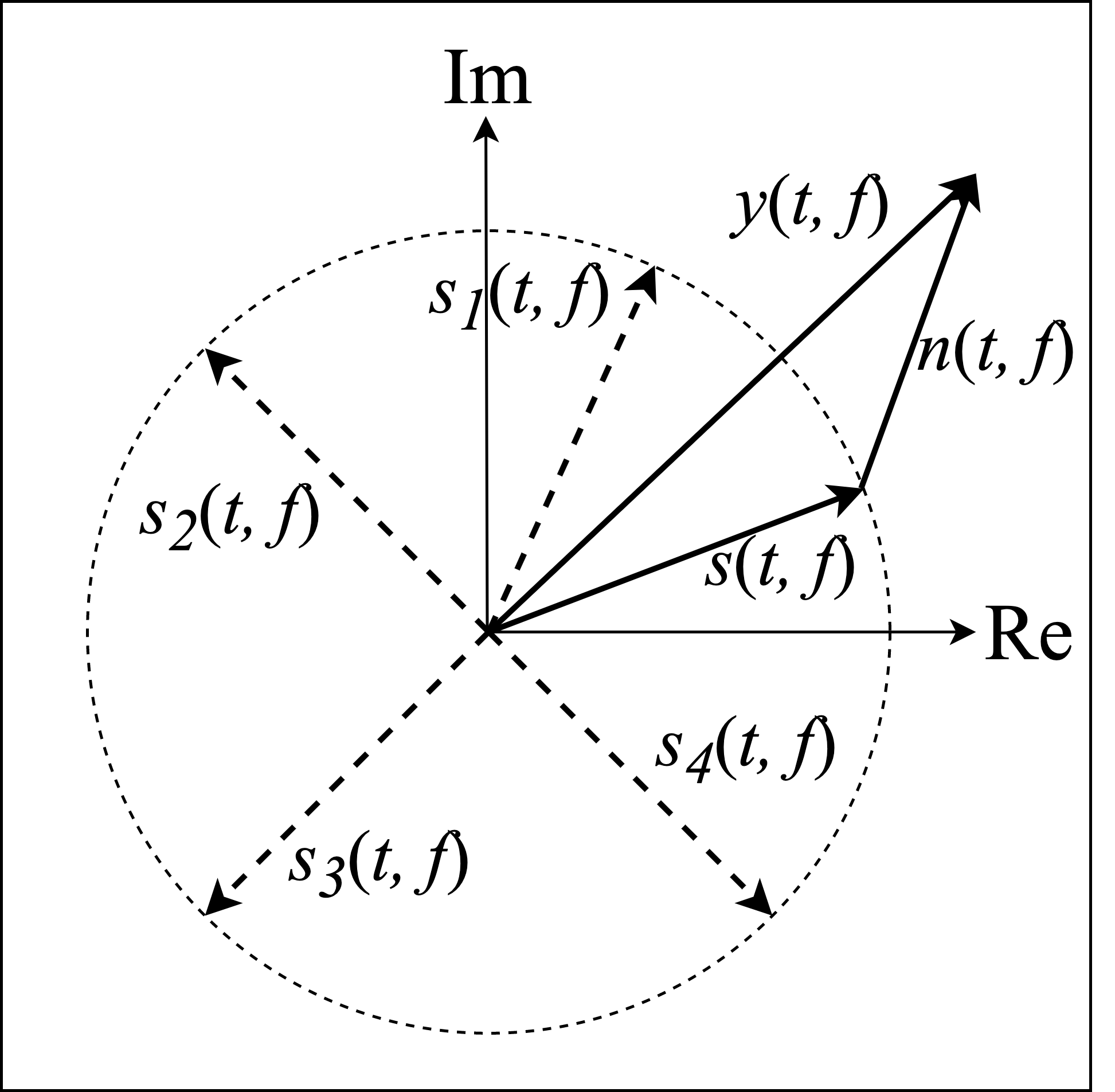}
\caption{$L_{SM}$}
\end{subfigure}%
\begin{subfigure}[h]{0.22\textwidth}
\centering
\includegraphics[width=0.98\textwidth]{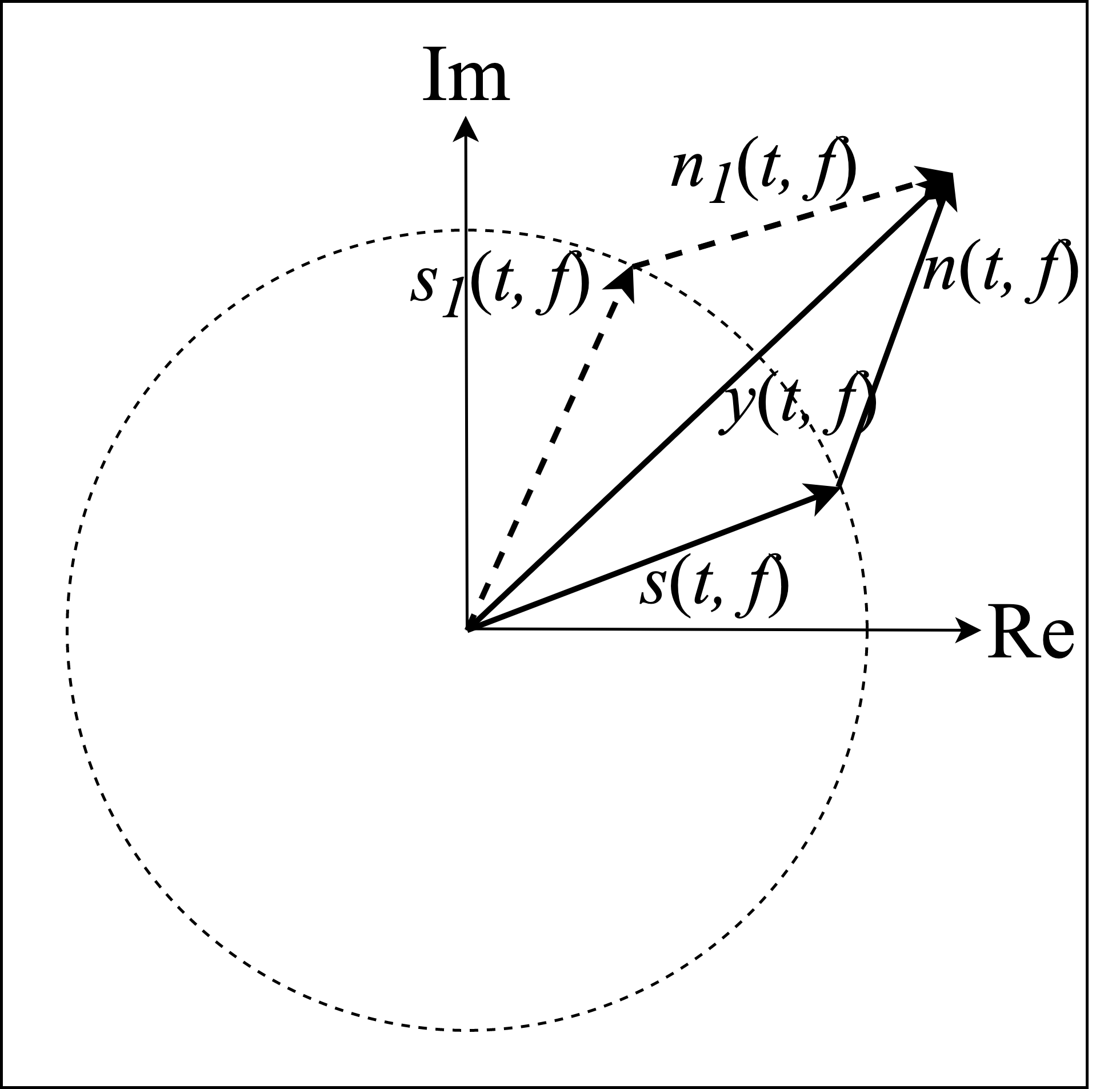}
\caption{$L_{PCM}$}
\end{subfigure}
\caption{Differences between $L_{SM}$ and $L_{PCM}$ in Cartesian (rectangular) coordinates. Re and Im respectively denote the real and the imaginary axes in complex plane.}
\label{fig_re_im}
\end{figure}

How can $L_{PCM}$ remove the artifact caused by $L_{SM}$? Let $y(t, f)$, $s(t, f)$, and $n(t, f)$ respectively denote the STFT coefficients at a given T-F unit of noisy speech, clean speech, and noise.  $L_{SM}$ aims at obtaining close estimates of $|s(t, f)|$ only, and there is an arbitrary number of perfect estimates of $|s(t, f)|$ in the complex representation. This is illustrated in Fig. \ref{fig_re_im}(a) with $5$ perfect estimates of $|s(t, f)|$ at the perimeter of a circle with the radius of $|s(t, f)|$. $L_{PCM}$, on the other hand, aims at getting good estimates of both $|s(t, f)|$ and $|n(t, f)|$, and it has only two candidates for the perfect estimate as shown in Fig. \ref{fig_re_im}(b). This implies that $L_{PCM}$ optimizes $L_{SM}$ with an additional constraint on phase, hence the name PCM. 

\section{Experimental settings}
\label{sec_experiments}
\subsection{Datasets}
We evaluate all the models in a speaker- and noise-independent way on the WSJ0 SI-84 dataset (WSJ) \cite{paul1992design}, which consists of $7138$ utterances from $83$ speakers ($42$ males and $41$ females). Seventy seven speakers are used for training and remaining six are used for evaluation. For training, we use 10000 non-speech sounds from a
sound effect library (available at \url{www.sound-ideas.com}) \cite{chen2016large}, and generate $320000$ noisy utterances at SNRs uniformly sampled from \{$-5$ dB, $-4$ dB, $-3$ dB, $-2$ dB, $-1$ dB, $0$ dB\}. 
For the test set, we use babble and cafeteria noises from an Auditec CD (available at \url{http://www.auditec.com}), and generate $150$ noisy utterances for both the noises at SNRs of $-5$ dB, $0$ dB, and $5$ dB.  

\subsection{System Setup}
All the utterances are resampled to $16$ kHz. We use $L = 512, J = 256, C = 64, E = 5,$ and $F = 32$. Inside a dense block, $m$ is set to $2$ for causal and $3$ for non-causal DCN. 

The Adam optimizer \cite{kingma2014adam} is used for SGD (stochastic gradient descent) based optimization with a batch size of  $4$ utterances.  All the models are trained for 15 epochs using a learning rate schedule given in \cite{pandey2020densely}. We use PyTorch \cite{paszke2017automatic} to develop all the models, and utilize its default settings for initialization.
DCN and NC-DCN are trained using two NVIDIA Volta V100 16GB GPUs and require one week of training. The DataParallel module of PyTorch is used to distribute data to two GPUs.

\subsection{Baseline Models}
We compare DCN with different existing approaches to speech enhancement, namely T-F masking, spectral mapping, complex spectral mapping, and time-domain enhancement. For T-F masking, we train an IRM based 4-layered bidirectional long short-term memory (BLSTM) network \cite{chen2017long}. A gated residual network (GRN) proposed in \cite{tan2018gated1} is used for spectral mapping. For complex spectral mapping, we report results from a recently proposed state-of-the-art gated convolutional recurrent network (GCRN) \cite{tan2019learning}. We compare with both causal and non-causal GCRN. For time-domain enhancement, we compare results with three different models: auto-encoder CNN (AECNN) \cite{pandey2019new}, temporal convolutional neural network (TCNN) \cite{pandey2019tcnn}, and speech enhancement generative adversarial network (SEGAN) \cite{pascual2017segan}. SEGAN is trained with the time-domain loss as we find it to be superior to adversarial training proposed in the original paper. 
\subsection{Evaluation Metrics}
We use short-time objective intelligibility (STOI) \cite{taal2011algorithm}, perceptual evaluation of speech quality (PESQ) \cite{rix2001perceptual}, and signal-to-noise ratio (SNR) as the evaluation metrics, which are the standard metrics for speech enhancement. STOI values typically range from $0$ to $1$, which can be roughly interpreted
as percent correct. PESQ values range from $-0.5$ to $4.5$.

\begin{table*}[!t]
\centering
\caption{Performance comparisons between different configurations of dense block, dilation, and attention in DCN. Boldface indicates the best score in a given condition.}
\label{tbl_ablation}
\centering
\begin{adjustbox}{width=0.86\textwidth}
\begin{tabular}{|c|c|c|c|cccc|cccc|cccc|cccc|cccc|cccc|}
\hline
\multicolumn{4}{|c|}{ Metric } & \multicolumn{8}{c|}{ STOI } & \multicolumn{8}{c|}{ PESQ } & \multicolumn{8}{c|}{ SNR } \\
\hline
\multicolumn{4}{|c|}{ Test noise } & \multicolumn{4}{c|}{ Babble } & \multicolumn{4}{c|}{ Cafeteria } & \multicolumn{4}{c|}{ Babble } & \multicolumn{4}{c|}{ Cafeteria } & \multicolumn{4}{c|}{ Babble } & \multicolumn{4}{c|}{ Cafeteria } \\
\hline
\multicolumn{4}{|c|}{ Test SNR (dB)} & -5 & 0 & 5 & Avg. & -5 & 0 & 5 & Avg. & -5 & 0 & 5 & Avg. & -5 & 0 & 5 & Avg. & -5 & 0 & 5 & Avg. & -5 & 0 & 5 & Avg. \\
\hline
\multicolumn{1}{|c|}{ Mix. $\rightarrow$} & \multicolumn{1}{c|}{ $m \downarrow$} & \multicolumn{1}{c|}{ Dil. $\downarrow$ } & \multicolumn{1}{c|}{ Att. $\downarrow$ } & 58.4 & 70.5 & 81.3 & 70.1 & 57.1 & 69.7 & 81.0 & 69.2 & 1.56 & 1.82 & 2.12 & 1.83 & 1.46 & 1.77 & 2.12 & 1.78 & -5.0 & 0.0 & 5.0 & 0 & -5.0 & 0.0 & 5.0 & 0.0 \\
\hline
\multirow{6}{*}{ \rotatebox{90}{Causal} } & 1 &  \xmark &  \xmark & 76.7 & 88.0 & 93.2 & 86.0 & 76.4 & 87.8 & 92.9 & 85.7 & 1.90 & 2.39 & 2.76 & 2.35 & 2.02 & 2.49 & 2.84 & 2.45 & 5.5 & 9.9 & 13.4 & 9.6 & 6.5 & 10.4 & 13.4 & 10.1 \\
& 2 & \xmark & \xmark & 81.6 & 91.3 & 95.0 & 89.3 & 80.5 & 90.2 & 94.3 & 88.3 & 2.13 & 2.70 & 3.08 & 2.64 & 2.17 & 2.68 & 3.05 & 2.63 & 7.4 & 11.5 & 14.7 & 11.2 & 7.7 & 11.4 & 14.4 & 11.2 \\
& 2 & \cmark &  \xmark & 83.5 & 91.9 & 95.2 & 90.2 & 81.4 & 90.5 & 94.5 & 88.8 & 2.23 & 2.75 & 3.12 & 2.70 & 2.21 & 2.70 & 3.07 & 2.66 & 7.7 & 11.8 & 15.0 & 11.5 & 7.9 & 11.5 & 14.5 & 11.3 \\
& 2 & \cmark & \cmark & 84.9 & 92.2 & 95.3 & 90.8 & 82.1 & 90.7 & 94.6 & 89.1 & 2.30 & 2.77 & 3.14 & 2.74 & 2.23 & 2.71 & 3.08 & 2.67 & 8.2 & 12.0 & \textbf{15.1} & 11.8 & \textbf{8.2} & \textbf{11.7} & \textbf{14.7} & \textbf{11.5} \\
& 2 &  \xmark & \cmark & \textbf{85.3} & \textbf{92.3} & \textbf{95.4} & \textbf{91.0} & \textbf{82.3} & \textbf{90.8} & \textbf{94.7} & \textbf{89.3} & \textbf{2.34} & \textbf{2.81} & \textbf{3.17} & \textbf{2.77} & \textbf{2.24} & \textbf{2.72} & \textbf{3.09} & \textbf{2.68} & \textbf{8.5} & \textbf{12.1} & \textbf{15.1} & \textbf{11.9} & \textbf{8.2} & \textbf{11.7} & \textbf{14.7} & \textbf{11.5} \\
& 1 &  \xmark & \cmark & 83.9 & 91.8 & 95.2 & 90.3 & 81.0 & 90.3 & 94.5 & 88.6 & 2.23 & 2.72 & 3.09 & 2.68 & 2.15 & 2.62 & 3.01 & 2.59 & 7.9 & 11.8 & 15.0 & 11.6 & 7.9 & 11.5 & 14.5 & 11.3 \\
\hline
\multirow{5}{*}{\rotatebox{90}{Non-causal} } & 3 &  \xmark &  \xmark & 84.7 & 92.5 & 95.7 & 90.9 & 83.1 & 91.4 & 95.0 & 89.8 & 2.37 & 2.88 & 3.22 & 2.82 & 2.34 & 2.82 & 3.16 & 2.77 & 8.2 & 12.2 & 15.2 & 11.9 & 8.3 & 11.8 & 14.7 & 11.6 \\
& 3 & \cmark &  \xmark & 86.6 & 92.9 & 95.7 & 91.7 & 84.1 & 91.7 & 95.0 & 90.3 & 2.53 & 2.96 & 3.24 & 2.91 & 2.44 & 2.88 & 3.19 & 2.84 & 9.1 & 12.5 & 15.3 & 12.3 & 8.7 & 12.0 & 14.8 & 11.8 \\
& 3 & \cmark & \cmark & \textbf{87.9} & \textbf{93.5} & 96.0 & 92.4 & 85.0 & 92.0 & 95.2 & 90.8 & \textbf{2.61} & 3.02 & 3.32 & 2.98 & \textbf{2.47} & \textbf{2.91} & \textbf{3.24} & \textbf{2.87} & \textbf{9.6} & \textbf{12.9} & 15.7 & 12.7 & \textbf{8.9} & 12.2 & 15.0 & 12.0 \\
& 3 &  \xmark & \cmark & \textbf{87.9} & \textbf{93.5} & \textbf{96.1} & \textbf{92.5} & \textbf{85.0} & \textbf{92.1} & \textbf{95.3} & \textbf{90.8} & \textbf{2.61} & \textbf{3.04} & \textbf{3.33} & \textbf{2.99} & 2.45 & \textbf{2.91} & 3.23 & 2.86 & \textbf{9.6} & \textbf{12.9} & \textbf{15.8} & \textbf{12.8} & \textbf{8.9} & \textbf{12.3} & \textbf{15.1} & \textbf{12.1} \\
& 1 &  \xmark & \cmark & 83.7 & 91.5 & 95.2 & 90.1 & 80.1 & 89.8 & 94.3 & 88.1 & 2.24 & 2.71 & 3.09 & 2.68 & 2.13 & 2.59 & 2.98 & 2.57 & 8.3 & 12.0 & 15.2 & 11.8 & 7.8 & 11.4 & 14.6 & 11.3 \\
\hline
\end{tabular}
\end{adjustbox}
\label{tbl:5}
\end{table*}

\begin{table*}[t]
\centering
\caption{STOI and PESQ comparisons between DCN and the baseline models of a) T-F masking, b) spectral mapping, c) complex-spectral mapping, and d) time-domain enhancement. }
\label{tbl_baselines}
\centering
\begin{adjustbox}{width=0.72\textwidth}
\begin{tabular}{|c|c|c|c|cccc|cccc|cccc|cccc|}
\hline
\multirow{4}{*}{ \rotatebox{90}{Approach}} & \multirow{4}{*}{  \rotatebox{90}{Causal?}} & \multirow{4}{*}{ \rotatebox{90}{Real-time?} } & Metric & & \multicolumn{7}{c|}{ STOI } & \multicolumn{8}{c|}{ PESQ } \\
\cline{4-20}
& & & Test Noise & \multicolumn{4}{c|}{ Babble } & \multicolumn{4}{c|}{ Cafeteria } & \multicolumn{4}{c|}{ Babble } & \multicolumn{4}{c|}{ Cafeteria } \\
\cline{4-20}
& & & Test SNR & -5 db & 0 dB & 5 dB & AVG & -5 dB & 0 dB & 5 dB & AVG & -5 db & 0 dB & 5 dB & AVG & -5 dB & 0 dB & 5 dB & AVG \\
\cline{4-20}
& & & Mixture & 58.4 & 70.5 & 81.3 & 70.1 & 57.1 & 69.7 & 81.0 & 69.2 & 1.56 & 1.82 & 2.12 & 1.83 & 1.46 & 1.77 & 2.12 & 1.78 \\
\hline
\hline
a) & \xmark & \xmark & BLSTM \cite{chen2017long} & 77.4 & 85.8 & 91.0 & 84.7 & 76.1 & 84.7 & 90.5 & 83.7 & 1.97 & 2.37 & 2.69 & 2.34 & 2.01 & 2.38 & 2.51 & 2.30 \\
\hline
b) & \xmark & \xmark & GRN \cite{tan2018gated1} & 80.2 & 88.9 & 93.4 & 87.5 & 79.4 & 88.0 & 92.9 & 86.8 & 2.16 & 2.63 & 2.97 & 2.59 & 2.23 & 2.62 & 2.96 & 2.60 \\
\hline
\multirow{2}{*}{ c) } & \cmark & \cmark & GCRN \cite{tan2019learning} & 82.4 & 90.9 & 94.8 & 89.4 & 79.1 & 89.3 & 94.0 & 87.5 & 2.17 & 2.70 & 3.07 & 2.65 & 2.10 & 2.60 & 2.99 & 2.56 \\
& \xmark & \xmark & NC-GCRN \cite{tan2019learning} & 87.0 & 93.0 & 95.6 & 91.9 & 84.1 & 91.7 & 95.1 & 90.3 & 2.53 & 2.96 & 3.25 & 2.91 & 2.40 & 2.85 & 3.17 & 2.81 \\
\hline
\multirow{9}{*}{ d) } & \cmark & \xmark & SEGAN-T \cite{pascual2017segan}& 81.5 & 90.3 & 94.1 & 88.6 & 79.8 & 89.5 & 93.5 & 87.6 & 2.11 & 2.62 & 2.97 & 2.57 & 2.15 & 2.61 & 2.94 & 2.57 \\
& \cmark & \xmark & AECNN-SM \cite{pandey2019new}& 82.6 & 91.5 & 95.1 & 89.7 & 81.1 & 90.7 & 94.5 & 88.8 & 2.21 & 2.80 & 3.17 & 2.73 & 2.23 & 2.76 & 3.12 & 2.70 \\
& \cmark & \cmark & TCNN \cite{pandey2019tcnn} & 82.8 & 91.3 & 94.8 & 89.6 & 80.6 & 89.8 & 94.0 & 88.1 & 2.18 & 2.70 & 3.06 & 2.65 & 2.14 & 2.62 & 2.98 & 2.58 \\
\cline{2-20}
& \cmark & \cmark & DCN-T & \textbf{85.3} & 92.3 & 95.4 & 91.0 & 82.3 & 90.8 & 94.7 & 89.3 & 2.34 & 2.81 & 3.17 & 2.77 & 2.24 & 2.72 & 3.09 & 2.68 \\
& \cmark & \cmark & DCN-SM & 85.2 & \textbf{92.7} & \textbf{95.8} & \textbf{91.2} & \textbf{82.5} & \textbf{91.3} & \textbf{95.1} & \textbf{89.6} & \textbf{2.35} & \textbf{2.93} & \textbf{3.31} & \textbf{2.86} & \textbf{2.33} & \textbf{2.85} & \textbf{3.22} & \textbf{2.80} \\
& \cmark & \cmark & DCN-PCM & 85.1 & \textbf{92.7} & \textbf{95.8} & \textbf{91.2} & \textbf{82.5} & \textbf{91.3} & \textbf{95.1} & \textbf{89.6} & 2.31 & 2.91 & 3.30 & 2.84 & 2.29 & 2.82 & \textbf{3.22} & 2.78 \\
\cline{2-20}
& \xmark & \xmark & NC-DCN-T & 87.9 & 93.5 & 96.1 & 92.5 & 85.0 & 92.1 & 95.3 & 90.8 & 2.61 & 3.04 & 3.33 & 2.99 & 2.45 & 2.91 & 3.23 & 2.86 \\
& \xmark & \xmark & NC-DCN-SM & \textbf{89.1} & 94.2 & 96.5 & \textbf{93.3} & \textbf{85.8} & 92.9 & 95.8 & \textbf{91.5} & \textbf{2.75} & \textbf{3.19} & 3.46 & \textbf{3.13} & \textbf{2.61} & \textbf{3.07} & 3.37 & \textbf{3.02} \\
& \xmark & \xmark & NC-DCN-PCM & 89.0 & \textbf{94.3} & \textbf{96.6} & \textbf{93.3} & 85.6 & \textbf{93.0} & \textbf{95.9} & \textbf{91.5} & 2.71 & 3.18 & \textbf{3.48} & 3.12 & 2.56 & 3.07 & \textbf{3.39} & 3.01 \\
\hline
\end{tabular}
\end{adjustbox}
\label{tbl:5}
\end{table*}

\section{Results and Discussions}
\label{sec_results}
\subsection{Ablation Study}
In this section, we present the findings of an ablation study performed to analyze the effectiveness of different context-aggregation techniques in DCN. There are $3$ components responsible for context-aggregation. First, using $m > 1$ in a dense block so that the receptive field of convolution extends beyond one frame. Second, using an exponentially increasing dilation rate in the layers of dense blocks, as proposed in \cite{pandey2020densely}. Third, the attention module proposed in this study (Section \ref{sec_attention}). STOI, PESQ, and SNR scores for causal and non-causal models trained using $L_{T}$ are given in Table \ref{tbl_ablation}. 

We observe that when there is no context, i.e., $m=1$, no dilation, and no attention, an average improvement of $16.2\%$ in STOI, $0.59$ in PESQ, and $9.9$ dB in SNR is obtained in causal enhancement. Increasing $m$ to $2$ with causal convolution obtains further improvement  of $3\%$ in STOI, $0.24$ in PESQ, and $1.3$ dB in SNR. Next, replacing causal convolutions with dilated and causal convolutions, as in \cite{pandey2020densely}, obtains further improvement of $0.7\%$ in STOI, $0.05$ in PESQ, and $0.2$ dB in SNR. Most of the improvements due to dilated convolutions are at the negative SNR of $-5$ dB. This suggests that a larger context is more helpful for speech enhancement in low SNR conditions. Further, inserting attention module to the network consistently improves objective scores with relatively larger improvements at $-5$ dB. In summary, objective scores are improved by progressively adding all the three components of context aggregation to the model, and most of the improvements are obtained at $-5$ dB.

Next, we change the dilated convolutions to normal convolutions and observe that objective scores either improve or remain similar. This suggests that using dilated convolutions along with attention would be redundant, since attention can utilize maximum available context. Thus we can expect that $m=1$ with attention should be sufficient for context aggregation. However, we find that reducing $m$ from $2$ to $1$ degrades performance. Therefore, context aggregation using the attention module along with some context with normal convolution is important for optimal results. Also, we find $m=3$ to be worse than $m=2$ (not reported here). A similar behavior is observed for non-causal models, where $m$ is set to $3$ instead of $2$ to maintain symmetry in context from past and future.

\subsection{Loss Comparisons}
\begin{figure}[!b]
\centering
\includegraphics[width=0.87\columnwidth, keepaspectratio]{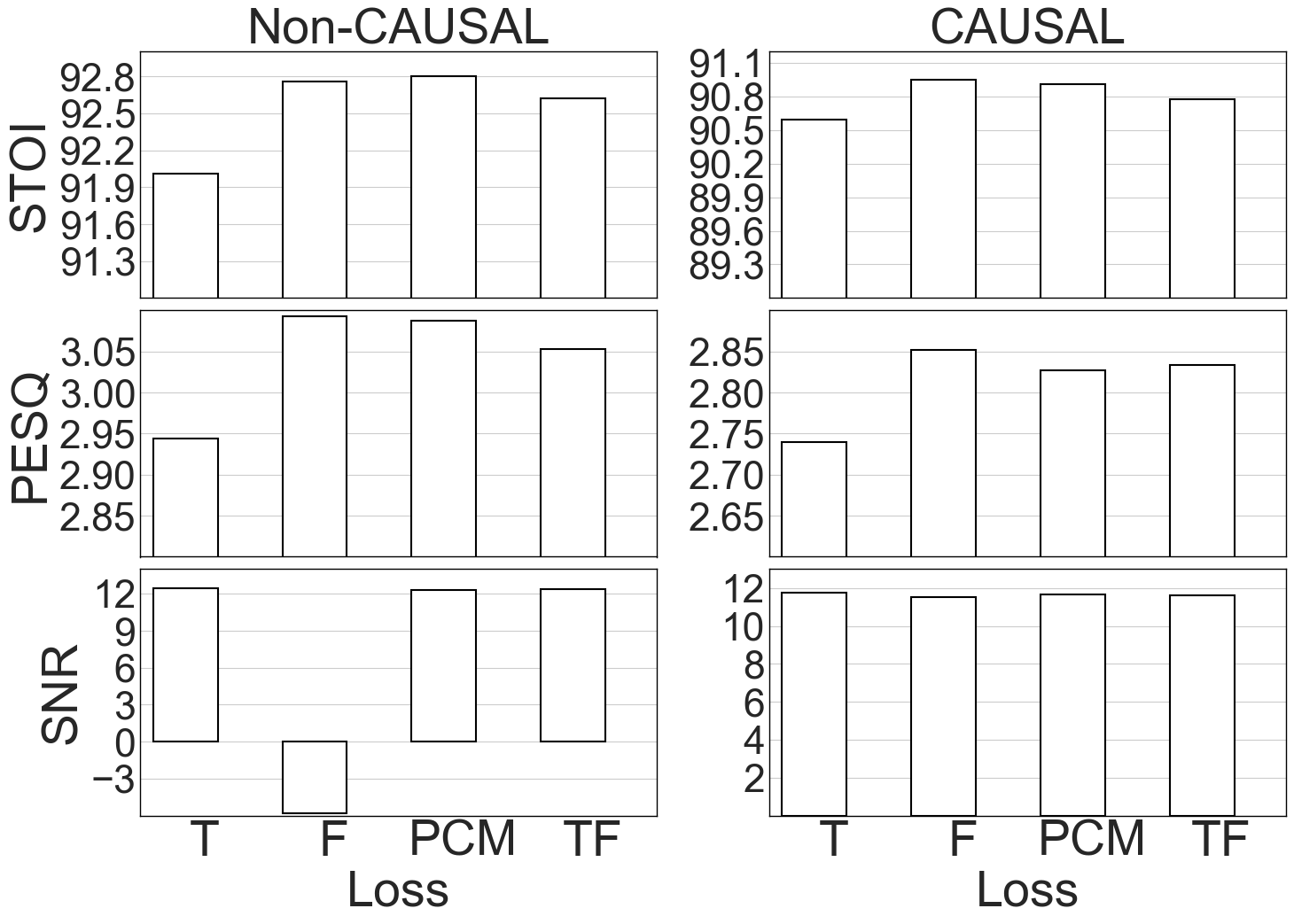}
\caption{STOI, PESQ and SNR comparisons between different loss functions.}
\label{fig_loss_pcm}
\end{figure}

This section analyzes different loss functions, and illustrates advantages of the proposed $L_{PCM}$. First, we reveal the inconsistent SNR improvement issue with $L_{SM}$. Causal and non-causal DCN are trained using $L_{T}$, $L_{SM}$, $L_{TF}$, and $L_{PCM}$, and average STOI and PESQ scores over two test noises and SNRs of $-5$ dB, $-2$ dB, $0$ dB, $2$ dB, and $5$ dB are plotted in Fig. \ref{fig_loss_pcm}. We observe that $L_{SM}$, $L_{TF}$, and $L_{PCM}$ obtain similar STOI scores, and they are better than $L_{T}$. $L_{T}$, $L_{TF}$, and $L_{PCM}$ obtain similar SNR scores, whereas $L_{SM}$ obtains similar SNR for a causal system but significantly worse SNR (even worse than unprocessed) for the non-causal system. We find that the SNR improvement of $L_{SM}$ is sensitive to learning rate, initialization and model architecture, i.e., not consistent. We also find that both $L_{TF}$ and $L_{PCM}$ obtain consistent SNR improvement similar to $L_{T}$, suggesting that $L_{TF}$ and $L_{PCM}$ can solve this issue without compromising STOI and PESQ scores.

Next, we evaluate the effects of $\alpha$ in $L_{TF}$. Average STOI, PESQ, and SNR scores of a dilation based model \cite{pandey2020densely} are plotted in Fig. 9 over two test noises and SNRs of $-5$ dB, $-2$ dB, $0$ dB, $2$ dB, and $5$ dB. We use $\alpha$ values from \{$0.0$, $0.2$, $0.4$, $0.6$, $0.8$, $1.0$\}. We can notice that for $\alpha < 1$, STOI and PESQ scores are similar. For $\alpha = 1$, which corresponds to $L_{T}$, STOI and PESQ results are worse. Similarly, SNR scores are similar for $\alpha > 0$ and worse for $\alpha = 0$, which corresponds to $L_{SM}$. These observations suggest that as long as $L_{SM}$ is included in training, better STOI and PESQ results are obtained. Similarly, as long as $L_{T}$ is included in training, a consistent improvement in SNR is obtained.

We provide enhanced speech samples at \url{https://web.cse.ohio-state.edu/~wang.77/pnl/demo/PandeyDCN.html}. The artifact is observed with $L_{SM}$ and $L_{TF}$, but not with $L_{T}$ and $L_{PCM}$. These comparisons suggest that $L_{TF}$ can solve the inconsistent SNR issue, but is not able to remove the artifact. Fig. 8 suggests that the proposed $L_{PCM}$ improves SNR consistently and obtains STOI and PESQ similar to $L_{SM}$. As shown in Fig. 6, the PCM loss removes the buzzing artifact present in the SM and TF losses.  

\begin{figure}[!b]
\centering
\includegraphics[width=\columnwidth, keepaspectratio]{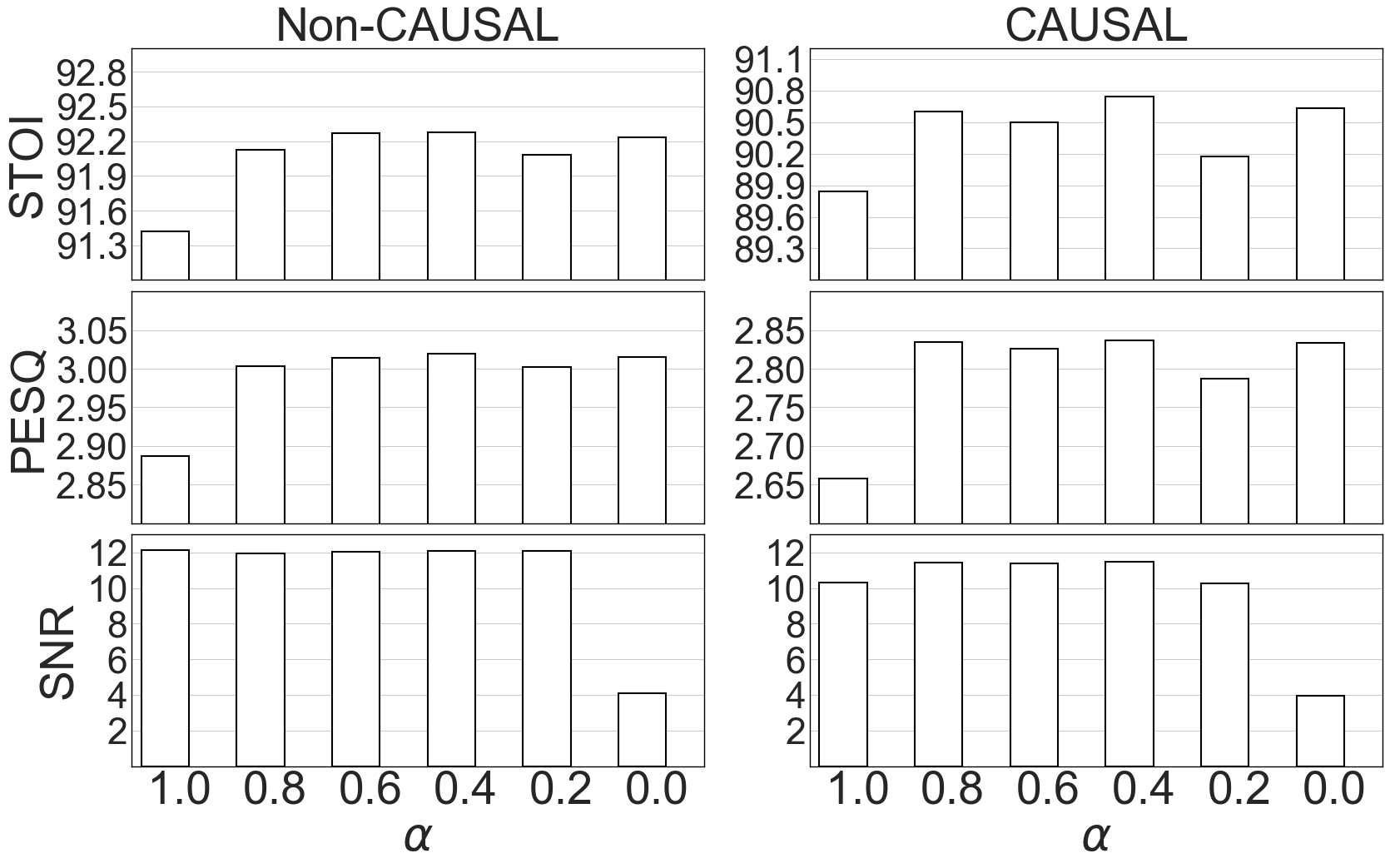}
\caption{Performance of $L_{TF}$ with different $\alpha$ values. }
\label{fig_alpha}
\end{figure}

\subsection{Comparison with Baselines}
In this section, we present results to demonstrate the superiority of DCN over different approaches. DCN is compared with a BLSTM for T-F masking \cite{chen2017long}, GRN \cite{tan2018gated1} for spectral mapping, GCRN \cite{tan2019learning} for complex spectral mapping, and SEGAN \cite{pascual2017segan}, AECNN \cite{pandey2019new}, and TCNN for time-domain enhancement. In our results, we call a system real-time if it is causal and uses a frame size less than or equal to $32$ ms, which is a general setting for real-time enhancement algorithms. The STOI and PESQ scores over two test noises are given in Table \ref{tbl_baselines}. We denote non-causal DCN as NC-DCN and non-causal GCRN as NC-GCRN. DCN trained with $L_{X}$ is denoted as DCN-X.

First, we observe that a frame based model with $m = 1$, no dilation, and no attention (Table \ref{tbl_ablation}), outperforms BLSTM based T-F masking on average. BLSTM is slightly better at $-5$ dB SNR. Note that BLSTM is a non-causal system that utilizes a whole utterance for one frame enhancement. This suggests that, even without any context information, the proposed model is a highly effective network for speech enhancement in the time domain. 

Further, using $m=2$ with causal convolution makes it significantly better than spectral mapping based non-causal GRN and time-domain SEGAN, which is a causal network but uses a frame size of $1$ second, and hence is not real-time. It is also similar or better than complex-spectral mapping based causal GCRN for all cases but babble noise at $-5$ dB. Similarly, using $m=3$ with non-causal convolution makes it comparable to NC-GCRN, which is the best performing network in the baseline models. It implies that the proposed network can outperform all the baselines without any dilation and attention. Also, these comparisons are done with the proposed network trained with $L_{T}$; training with $L_{PCM}$ will obtain obtain even better performance improvement over baselines. 

Additionally, Table II reports STOI and PESQ numbers for DCN-T, DCN-SM, and DCN-PCM. We can see that DCN-SM and DCN-PCM obtain similar scores, which are better than DCN-T for all the cases except babble $-5$ dB, where scores are similar for all the three losses. 

Finally, we compare DCN-PCM, the best real-time version, with other real-time baselines. For real-time systems, TCNN is the best baseline, and DCN, on average, is better than TCNN by $1.5\%$ for STOI and $0.19$ for PESQ. Similarly, we compare NC-DCN-PCM with NC-GCRN, the best non-causal baseline system. NC-DCN, on average, outperforms NC-GCRN by $1.3\%$ in STOI and $0.21$ in PESQ. The $p$ values for statistical significance test between GCRN and DCN-PCM, and between NC-GCRN and NC-DCN-PCM are found to be less than 0.0001  at all SNRs for both STOI and PESQ, indicating statistically significant improvements.

\subsection{Attention Maps}

\begin{figure}[h]
\centering
\includegraphics[width=0.45\textwidth]{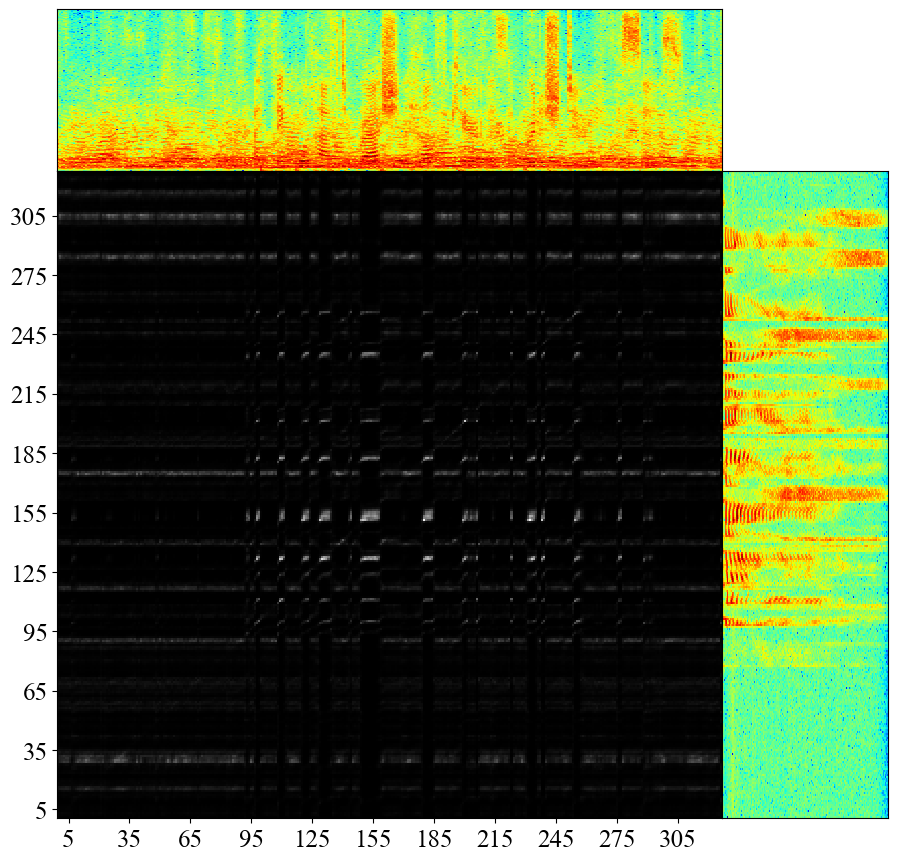}
\caption{Attention map of a sample utterance with non-causal DCN.}
\label{fig_nc_att}
\end{figure}

\begin{figure}[h]
\centering
\includegraphics[width=0.45\textwidth]{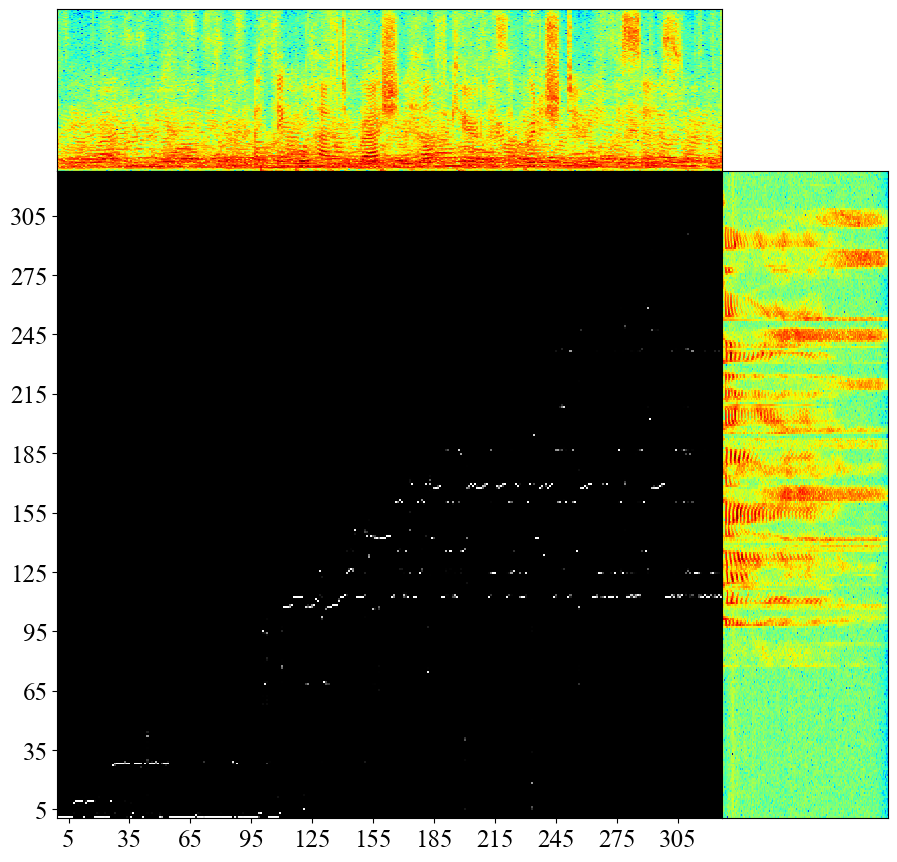}
\caption{Attention map of the same utterance as in Fig. 10 with causal DCN.}
\label{fig_att}
\end{figure}

The attention mechanism in DCN is meant to focus on the frames of an utterance that can aid speech enhancement. In this section, we plot attention scores of Eq. (13) for non-causal and causal DCN. Attention scores for a sample utterance from the last layer of the encoder of DCN are plotted in Fig. \ref{fig_nc_att} and Fig. \ref{fig_att}. The horizontal axis represents the frame index of interest, and the vertical axis represents the frames over which a given frame attends to. The spectrogram on top shows the noisy speech and the one on the right clean one.

For non-causal DCN, we observe that the most of the attention is paid to the harmonic structure, i.e., on voiced speech, between frames $125$ and $185$. Also, there is some attention to two high-frequency sounds towards the end of the utterance.  

For causal DCN, since frames in the future are not available, the attention on voiced sounds has shifted to earlier frames above frame $95$. For high-frequency sounds, the two sounds towards the end of the utterances that are used in non-causal case are not available, and hence the attention is shifted to earlier high-frequency sounds between frames $155$ and $185$. Also, attention in causal DCN is sharper than that in non-casual DCN.

\section{Concluding Remarks}
In this study, we have proposed a novel dense convolutional network with self-attention for speech enhancement in the time domain. The proposed DCN is based on an encoder-decoder structure with skip connections. The encoder and decoder each consists of dense blocks and attention modules that enhance feature extraction using a combination of feature reuse, increased depth, and maximum context aggregation. We have evaluated different configurations of DCN, and found that the attention mechanism in conjunction with a normal convolution with a small receptive field, i.e, no dilation, is helpful for time-domain enhancement. We have developed causal and non-causal DCN, and have shown that DCN substantially outperforms existing approaches to talker- and noise-independent speech enhancement.

We have revealed some of the existing problems with a spectral magnitude based loss. Even though magnitude based loss obtains better objective intelligibility and quality scores, it is inconsistent in terms of SNR improvement, and introduces an unknown artifact in enhanced utterances. We have proposed a new phase constrained magnitude loss that combines the two losses over STFT magnitudes of the enhanced speech and predicted noise. The PCM loss solves the SNR and artifact issues while maintaining the improvements in objective scores.

By visualizing attention maps, we have found that most of the attention seems to be paid to voiced segments and some high-frequency regions. Further, attended regions appear different for causal and non-causal DCN, and attention is relatively sharper for causal speech enhancement. 

DCN is trained on the WSJ corpus and evaluated on untrained WSJ speakers. We have recently revealed that DNN-based speech enhancement fails to generalize to untrained corpora, and better performance on a trained corpus does not necessarily lead to a better performance on untrained corpora \cite{pandey2020cross, pandey2020learning}. For future research, we plan to evaluate DCN on untrained corpora, and explore techniques to improve cross-corpus generalization.

\label{sec_conclusions}

\bibliographystyle{IEEEtran}
\bibliography{IEEEabrv, mybib}
\end{document}